\documentclass[12pt,halfline,a4paper]{ouparticle}
\usepackage{listings}
\usepackage{hanging}
\usepackage{color, colortbl}
\definecolor{Gray}{gray}{0.9}
\usepackage{endnotes}
\let\footnote=\endnote

\begin{document}

\title{The Foundations of Political Realism}

\author{%
\thanks{The author thanks Robert Axelrod, Les Birnbaum, Sunanda Ghosh, Kenan Huremovic, and Nick Misera for their feedback.}
\name{Michael Poulshock}
\address{Drexel University, Thomas R. Kline School of Law}
}

\abstract{Political realism aims to describe the interaction of agents involved in struggles for political power. This article formulates realism in terms of quantitative postulates that depict political power as a fluid-like substance flowing through a network of agents. It is shown that agent preferences within this framework resemble those of nation states competing for power. It is suggested that this approach provides a methodology for resolving longstanding debates about realism and a template for interpreting aspects of political behavior more generally.}

\maketitle

\section{Introduction}

At its essence, political realism is the idea that the pursuit of power is the central feature of political activity. From the time this notion was first articulated by Thucydides almost 2,500 years ago, it has offered, to a first approximation, the strongest general account of political behavior yet devised. Over the centuries, political realism has developed from a loose set of historical observations, precepts, and claims about human nature (e.g. Morgenthau 1954; Machiavelli 2007) into a social science (see Glazer 2010; Tellis 1995; Jervis 1994; Keohane 1986; Waltz 1979; Kaplan 1957). The objective of this article is to advance that effort by putting political realism on a quantitative theoretical foundation.

Political realism posits that when agents interact under anarchic conditions, they are compelled to pursue power in order to promote their own security.\footnote{ Descriptive and prescriptive claims are often conflated in debates about realism. Our approach is descriptive and avoids normative statements about how political agents should behave.} The most sophisticated expression of political realism can be found in the field of international relations, particularly as elaborated by the theory of the balance of power. This theory, of which there are many variants (Sheehan 1995), attempts to describe the behavior of nation-states in an international system characterized by the absence of any supranational authority. Balance of power theory claims that states act to counterbalance powerful rivals to prevent them from dominating the system, and it attempts to evaluate the stability of unipolar, bipolar, and multipolar configurations of states, to determine which arrangements are more prone to violent conflict.

Despite its flagship status, there is room for improvement in balance of power theory. First, some of the theory's specific predictions do not align with the empirical evidence (Wohlforth 2007; Schroeder 1994). Second, the theory is based upon an unrepresentative sample of events: generally, modern European history (Buzan 2010). Third, there's no obvious way to choose among competing realist theories, such as balance of power, hegemonic stability, and power transition theories. Finally, the theory's predictions do not follow inevitably from an underlying set of fundamental principles, that is, from an axiomatic structure describing the phenomenon of political power. For example, scholars continue to debate what states' goals are, how power is pursued, and how balancing works (Vasquez 2003). A quantitative theory, in contrast, would produce objectively derivable predictions (c.f. Stoll 1987). Setting political realism on an axiomatic foundation has the potential to solidify balance of power theory and provide a generalization applicable to a wider spectrum of political phenomena.

This paper offers a quantitative theoretical foundation for political realism based on the notion that power is a fluid-like substance that flows through a network of agents. Agents can use their power to make others more powerful or, conversely, to diminish their power. In choosing how to deploy their power, agents alter the course of its movement. Consequently, power is both a stock (a quantity possessed by an agent) and a flow (a quantity in motion between agents). We call this approach \textbf{quantitative realism}, to distinguish it from realism as traditionally, qualitatively, conceived. Our hypothesis is that quantitative realism plausibly models the essence of political realism. (For other theories relating power and networks, see Huremovic 2014; Hafner-Burton 2009.) 

The approach presented here is necessarily abstract. It omits many variables that are relevant to political behavior, such as institutions, group identity, resource scarcity, technology, geography, norms, personality, and deception. Instead, it provides a parsimonious account of a single variable: power. This abstraction is useful because political realism is fundamentally about the dynamic nature of power struggles, and here we isolate that particular phenomenon in order to illuminate its mechanism (c.f. Goddard 2016; Axelrod 1995).

\section{The Foundations of Realism}

We first describe the foundations of political realism, starting from a small set of definitions and postulates and building up to a mathematical formalism. We then illustrate the ideas visually.\footnote{ This treatment is similar to Poulshock 2017.}

\subsection{Foundations}

\subsubsection{Definitions}

There are four relevant definitions. First, an \textbf{agent} is an individual actor or a collection of individuals presumed to be acting in concert, in other words, a political unit. Second, \textbf{power} is a quantity reflecting an agent's ability to affect the power of other agents. This is a deliberately recursive definition that strikes at the heart of what is obvious about power: those who have it can render others powerless, or conversely, empower them.\footnote{ The social sciences typically define power as the ability of one agent to compel another to do something it would not otherwise do. There is not necessarily a conflict between this traditional definition and the one proposed above, which generalizes the means of compulsion.} Third, an \textbf{action} is an allocation or transfer of power from one agent to another. Actions are what cause other agents' power levels to change. Finally, a \textbf{tactic} is an agent's foreign policy: a list of percentages indicating the agent's actions with respect to the other agents.

\subsubsection{Postulates}

The amount of power that agents possess is altered as a result of their interactions. There are six principles, or postulates, that quantify this process and that serve as the theoretical underpinning for political realism. These are:

\begin{enumerate}
  \item \textbf{Constructive action is an expenditure of power that increases the power of another agent by more than the amount expended.} Constructive action among states might entail trading goods or services, forming a military alliance, supporting an ally with troops and weapons, paying tribute, providing disaster relief, engaging in cultural exchange, or giving other forms of assistance. As a result of a constructive action, the acting agent loses power and the receiving agent gains power. For instance, in a transfer of one unit of constructive power, the acting agent will lose one unit and the receiving agent will gain more than one unit. When agents cooperate by reciprocating constructive exchanges, they make each other more powerful, with each becoming better off than they were before the exchange.
  \item \textbf{Destructive action is an expenditure of power that decreases the power of another agent, with more impact than constructive action.} Destructive action entails the use of violence or the imposition of unwanted consequences. It includes actions like military assault, siege, bombardment, killing, destruction of property and infrastructure, and terrorism. Both the acting agent and the receiving agent lose power as a result of destructive action.\footnote{ This mirrors Frederick Lanchester's linear law of the relationship between opposing military forces.} Since it is easier to harm another agent than to help them, and easier to destroy value than it is to create it, destructive action has a greater impact than constructive action. When a unit of destructive power is transferred, the acting agent loses one unit and the receiving agent loses more than they would have gained had the transfer been constructive.
  \item \textbf{Agents tend to dissipate power.} Agents tend to become weak. They need to consume some baseline amount of constructive power to survive, and if they don't, they will eventually perish. Power that is not actively used tends to depreciate.\footnote{ Experimentally, it is not clear that this postulate is necessary, but the assumption that strength remains static seems implausible.}
  \item \textbf{Agents prefer other agents to be relatively weak.} Agents are preoccupied with how much power other agents have. They want to be relatively powerful while keeping others relatively powerless. Moreover, given the choice, agents would prefer that their competition be weak and divided, rather than strong and united. For example, an agent would generally prefer five competitors with one unit of power each to one competitor with five units of power.\footnote{ By way of contrast, Mearsheimer (2014) argues that nation states seek to maximize their market share of total power. An agent using Mearsheimer's utility function would be unable to differentiate between these two scenarios.}
  \item \textbf{Agents want to become more powerful in absolute terms.} Agents want the amount of power they have to go up, rather than to decrease or remain constant.
  \item \textbf{Agents engage in ongoing interaction.} They do not interact merely one time. Nor do they necessarily know how long the interaction will continue, but they expect to continue interacting with each other in the future.
\end{enumerate}
These starting assumptions provide a framework for generating the behavior traditionally of concern to political realism.

We will refer to the relative strengths of the agents, together with their interrelationships, as a \textbf{power structure}, a term intended to align with conventional usage. Postulates 1 and 2 imply that when agents decide upon their foreign policy, they take both features of this object into consideration.\footnote{ One might generalize this by allowing agents to consider not just the present power structure but its history as well.} The first signal, relative strength, is referred to in the balance of power literature as the distribution of power, and is often thought as the market share of total power. The second signal is the intensity of the constructive and destructive actions among the agents. When an agent determines its foreign policy, a primary consideration is the power structure in which the agent is embedded.

Though on the surface it might appear that these starting assumptions are a radical departure from realism as traditionally conceived, they can be interpreted as a generalization of those of traditional realism. The traditional assumptions that states have military capacity and interact under anarchy are, here, abstracted into the simple idea that agents can use their power to render each other powerless, and possibly even nonexistent. And in both paradigms, agents interact repeatedly and are assumed to take self-help action to promote their own survival in light of the current distribution of power. What this new formulation adds are the following: First, the conception of power is more general. It's not solely about destructive capacity and it's not zero sum: power can be created as a result of cooperation. Second, agents make tactical decisions based upon their interrelationships and not solely upon the distribution of power, as in traditional realism. Finally, agents' use of power has quantitative effects on other agents' power levels, causing the distribution of power to change in potentially measurable ways. Overall, this generalization allows quantifiable consequences to follow directly from the starting assumptions.

\subsection{Formalization}

Readers who are not mathematically inclined can skip this section without loss of appreciation for the overall argument.

\subsubsection{Data Structures}

The principles above can be formalized as follows. The variable \textit{s}, for size, is a nonnegative number representing the amount of power an agent has. The vector \textbf{s} lists the sizes of all agents in the power structure. Each agent forms a tactic vector \( \boldsymbol{\tau} \) which indicates how that agent allocates its power towards the other agents (and itself). These allocations can be positive, indicating constructive action, or negative, indicating destruction. The tactic vector must satisfy the constraint
\begin{equation}
    \sum_{j=1}^{n} \left|\tau_{j}\right| = 1,  \hspace{1cm}  \tau_{j} \in [-1,1]
\end{equation}
where \textit{n} is the number of agents. That is, each of an agent's outgoing allocations \( \boldsymbol{\tau}_{j} \) is a percentage of the agent's total power, which must sum to 1. A weighted tactic matrix \textbf{T} is formed by assembling the individual agents' tactic vectors together as column vectors. Destructive actions are represented by negative values in \textbf{T}. The size vector and tactic matrix \{\textbf{s}, \textbf{T}\} together describe a power structure. Starting from these data structures, the first five postulates can be expressed mathematically.\footnote{ Postulate 6 is represented by time \textit{t}.}

\subsubsection{Law of Motion}

We can define a \textbf{law of motion} that describes how power flows through a network of agents. This equation describes how agent sizes change from one instant to the next, as power is transferred according to the tactic matrix. The formula follows directly (and trivially) from Postulates 1-3, and uses matrix multiplication to update the sizes of all agents at once:\footnote{ When an agent dies, its power cannot be less than 0. Any negative sizes that result from the law of motion must be set to 0, an operation not stated explicitly in this equation.}
\begin{equation}
    \mathbf{s}(t+1) = (\mathbf{T} \circ \mathbf{M}) \mathbf{s}(t)
\end{equation}
The \( \circ \) operator is the Hadamard product (element-wise matrix multiplication). \textbf{M} is a multiplier matrix defined as
\begin{equation}
    (m_{ij}) = \begin{cases} 
      \lambda & i = j \\
      \beta & \tau_{ij} \ge 0 \\
      \mu & \tau_{ij} < 0 \\
   \end{cases}
\end{equation}
with the parameters
\begin{equation}
\begin{split}
    \beta > 1 \\
    \mu > \beta \\
    \lambda \le 1
\end{split}
\end{equation}
We illustrate the law of motion in section 2.3.

\subsubsection{Utility Function}

Postulates 4 and 5 together describe what agents want: they want to be powerful in an absolute sense, and they want to dominate others by being powerful in a relative sense. These are sometimes in tension (Powell 1991), as when one agent can only grow in absolute terms by allowing another agent to grow even more, or when one agent can dominate another in a relative sense only by losing power. We can encode these conflicting objectives into a utility function:\footnote{ See Poulshock 2017 for a justification of this utility function.}
\begin{equation}
u_{i} = \frac{s_{i}^\alpha}{\displaystyle\sum_{j=1}^{n} s_{j}^2},\enspace \enspace \alpha > 2
\end{equation}
As the utility exponent \( \alpha \) decreases, relative power and therefore destructive action are incentivized; as \( \alpha \) increases, absolute growth, which requires cooperation, is incentivized.

\subsubsection{Tactics and Implementation-Specific Assumptions}

Agents consider both the sizes and tactics of the other agents when deciding what to do. Formally, the tactic vector is a function \textit{f} of the current power structure:
\begin{equation}
    \boldsymbol{\tau}_{t+1} = f(\boldsymbol{s}_{t},\boldsymbol{T}_{t})
\end{equation}
A critical task is to find the function \textit{f}, as this is what animates agent behavior. The law of motion and the equation above form an iterative, dynamic process in which s is a function of \textbf{T}, and \textbf{T} in turn is a function of \textbf{s}. Equation (2) describes how the capacities of the agents change as a result of their tactical configurations. Equation (6) says that agents change their tactics in response to the changes in the power structure. The interplay between the two drives the model. Finding candidates for the function \( f(\boldsymbol{s}, \boldsymbol{T}) \) would be an important step in understanding the implications of quantitative realism. However, we do not do this here (c.f. Bueno de Mesquita 2004; Axelrod 1993).

We make additional assumptions specific to this research. First, the law of motion is implemented using an additional parameter \( \rho \), which sets a fixed percentage of each agent's power that is always allocated to itself. This parameter controls the tempo of a simulation: the more power that agents allocate to themselves, the less they allocate to others, and therefore the less change there is from one time step to the next. Second, we assume that agents have complete information: they know the entire power structure.\footnote{ This is obviously a significant simplifying assumption, one that is almost never true in the real world, in which the management of information is the primary art of statecraft and politics. There, agents may not know the entire power structure with certainty. Their knowledge is approximate, gleaned from signals, and often subject to misinformation, deception, and cognitive errors. Nonetheless, in the international context, it is reasonable to assume that actors for the most part operate with a shared understanding of the salient power relationships.} Third, events proceed in discrete time.

\subsection{Visualization}

We can depict power structures as graphs in which nodes represent agents and edges signify the relationship between two agents. The larger a node is, the more powerful that agent is. Solid arrows represent constructive actions, and dashed arrows show destructive ones. When two agents have the same attitude toward each other, the relationship is displayed as an undirected edge (one without arrows). Asymmetrical relationships are displayed as directed edges. Node colors will be explained in Section 3.
\begin{figure}[H]
\centering
\includegraphics[scale=.5]{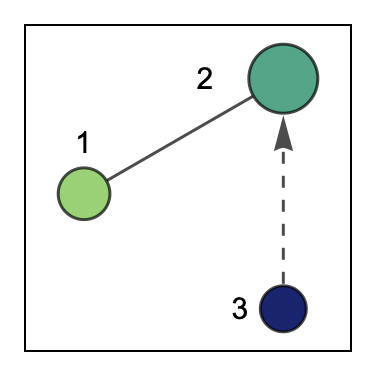}
\end{figure}
\noindent
The agents have an implicit numbering: the agent in the 9 o'clock position is number 1, and they increment going clockwise. In the figure above, agents 1 and 2 are cooperating, and agent 3 is attacking agent 2. Agent numbers are sometimes omitted from the diagrams below.

If someone were to look at a power structure diagram like the one above, without knowing any other historical or political details about a situation, they would have a decent first approximation of the power struggle at hand. When we add the dimension of time and apply the law of motion, power structures evolve in an intuitive fashion:
\begin{figure}[H]
\centering
\includegraphics[scale=.5]{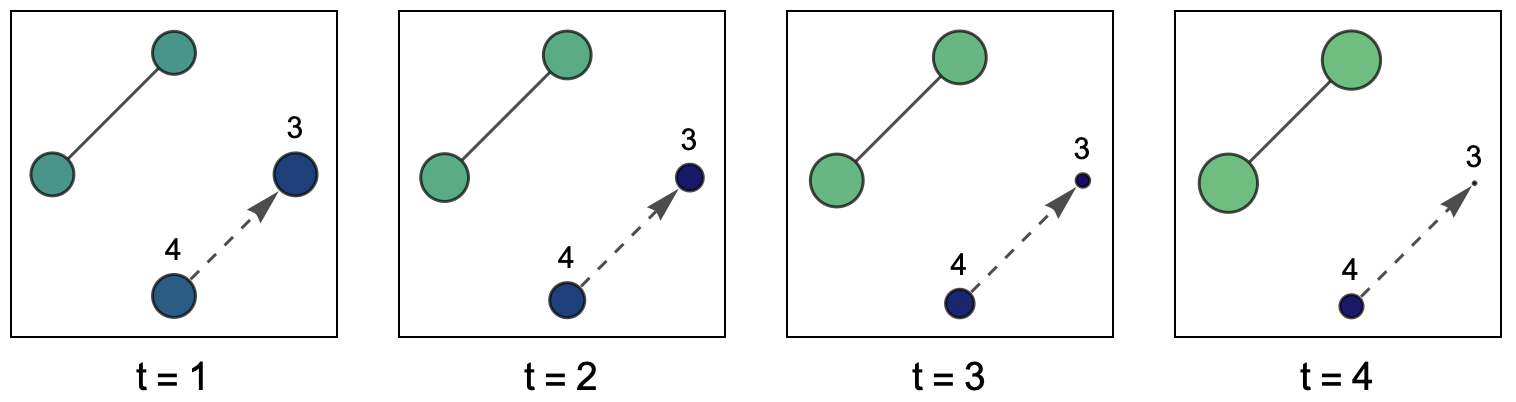}
\end{figure}
\noindent
The sequence above shows how agents' power levels change as a result of their interactions. Two of the agents are growing as a result of their cooperation, and agent 3 is shrinking due to agent 4's attack. Agent 4 is also shrinking, at a slower rate than 3, because it is expending power destroying agent 3. This sequence assumes that none of the agents alters their tactic vector. It illustrates the law of motion when T is held constant. Of course, nation-states don't keep their foreign policies constant; they adjust them to advance their interests within the power structure at hand. Nevertheless, the effect of the law of motion should be evident from this sequence.

\section{Methodology}

We explore the consequences of the six postulates to assess whether the model is a lucid expression of political realism. We  argue that agent preferences reflect the behavior of agents embroiled in real world power struggles.

By \textit{preference}, we mean a specific metric based on the following: Given a power structure, we apply the law of motion to it, calculate the agents' utility at each time step, discount future utility payoffs at a rate defined by the parameter \( \delta  \in (0, 1) \), and then sum. This is called intertemporal utility:
\begin{equation}
U_{i} = (1-\delta) \sum_{t=1}^{\infty} \delta^t u_{i}(t)
\end{equation}
This equation represents an agent's naive appraisal of utility as it accrues over time, naive in the sense that the agent assumes (implausibly) that all of the agents, including itself, will hold their tactics constant over those future time steps. The equation is an interpretation of Postulate 6 and a common method of discounting utility. With a sly wink at Machiavelli, we will refer to this metric \textit{U} as \textbf{PrinceRank}. It produces a complete ordering of all possible power structures, reflecting the preferences of a given agent. We use PrinceRank to understand which power structures are favorable to agents. Without loss of generality, we restrict our attention to discrete tactics in which agents can only select between positive, negative, and neutral actions.

We color the nodes of each graph along a blue-green-yellow spectrum to reflect the PrinceRank of each agent. 
\begin{figure}[H]
\centering
\includegraphics[scale=.5]{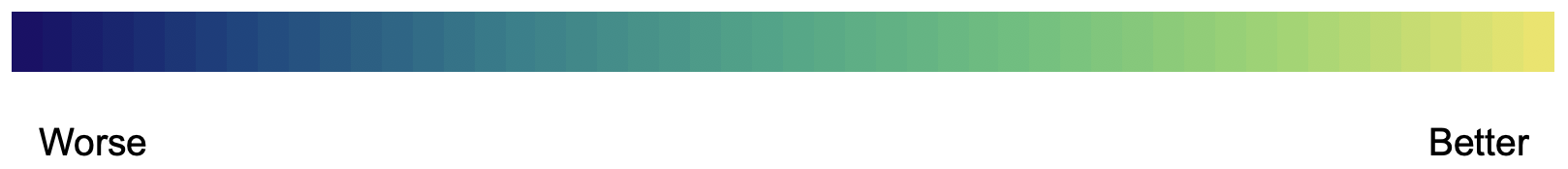}
\end{figure}
\noindent
Agents colored yellow have the highest PrinceRank, while blue agents have the lowest. This scheme allows the reader to see agent preferences at a glance. Note that in some figures, colors are exaggerated to distinguish subtle differences, so one cannot reliably compare a figure in one section with that in another.

Because we don't have a candidate function \textit{f}, we use two heuristics to guide our thinking about what actions an agent might take in a given situation. One is that asymmetrical actions are likely to be short-lived: constructive action will only be continued if it is reciprocated, and an agent who is attacked will attack back (Axelrod 1984). The second is that cooperation generally only happens when both agents benefit from it, whereas conflict can arise when a single agent has an incentive for it. 

Our use of PrinceRank has limitations. First, we do not completely take into account agent countermoves. The model posits indefinite interaction among the players, but most of our analysis is on a power structure at a single point in time. To analogize to the game of chess, it's as if we're assessing the value of a board position, without simulating a sequence of moves in the game tree. Second, there are six parameters that affect preferences, no combination of which is obviously correct. The simulations here use the parameter values in the rightmost column of the table below. These values were picked for no special reason other than that they seemed to give rise to the desired behavior.

\begin{figure}[H]
\centering
\begin{tabular}{ | c | c | c | c | }
\hline
\rowcolor{Gray}
\textbf{Symbol} & \textbf{Parameter} & \textbf{Range} & 
\textbf{Value Used} \\
\hline
$\beta$ & Constructive multiplier & $\beta > 1$ & 2 \\
$\mu$ & Destructive multiplier & $\mu > \beta$ & 3 \\
$\lambda$ & Decay multiplier & $\lambda \le 1$ & 1 \\
$\alpha$ & Utility exponent & $\alpha \ge 2$ & 2.25 \\
$\rho$ & Tempo control & $0 < \rho \le 1$ & 0.9 \\
$\delta$ & Discount rate & $0 < \delta < 1$ & 0.9 \\
\hline
\end{tabular}
\caption{Model parameters (the last two are extrinsic)}
\label{fig:variables}
\end{figure}

\section{Results and Discussion}

In this section, we show that agent preferences, as illuminated by PrinceRank (\textit{U}), generally align with those anticipated by political realism.

\subsection{Structural Ideals}

If an agent could structure the relations of a system of equally-sized agents however it wished, what would that power structure look like? When we explore possible tactic matrices and select the one that gives an agent the highest PrinceRank, we get something like the following:
\begin{figure}[H]
\centering
\includegraphics[scale=.5]{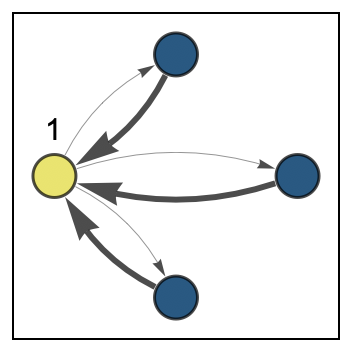}
\end{figure}
\noindent 
The agent 1's ideal arrangement is for every agent in the system to be giving it constructive power, and for it to be reciprocating to some lesser degree. When an agent is shortsighted (low $\delta$) or not interested in absolute growth (low $\alpha$), then its ideal structure is just to receive constructive power from every other agent and not reciprocate anything in return. However, forward-thinking agents reciprocate in order to create a positive feedback loop of mutual growth, albeit one based upon unequal exchange. If the hub-and-spoke structure above is every agent's fantasy, we would expect that when they have the ability to shape networks, they will try to impose this topology, which is reminiscent of imperial systems and their tributary states.

\subsection{Preferences in Triadic Structures}

We next focus our attention on triadic structures\footnote{ We omit discussion of two-agent power structures, which are conceptually similar to the iterated Prisoner's Dilemma. Axelrod 1984.} in which every relationship is reciprocal in polarity and in which all agents have the same size. There are 18 such scenarios that are unique from agent 1's perspective. When we sort these by PrinceRank, there is a fair amount of variation in the ordering that results from the parameters chosen. However, the most preferred and least preferred structures tend to remain in the same general place in the list, regardless of the parameter choices. Here are agent 1's four most preferred structures, under the parameters used throughout this paper:
\begin{figure}[H]
\centering
\includegraphics[scale=.5]{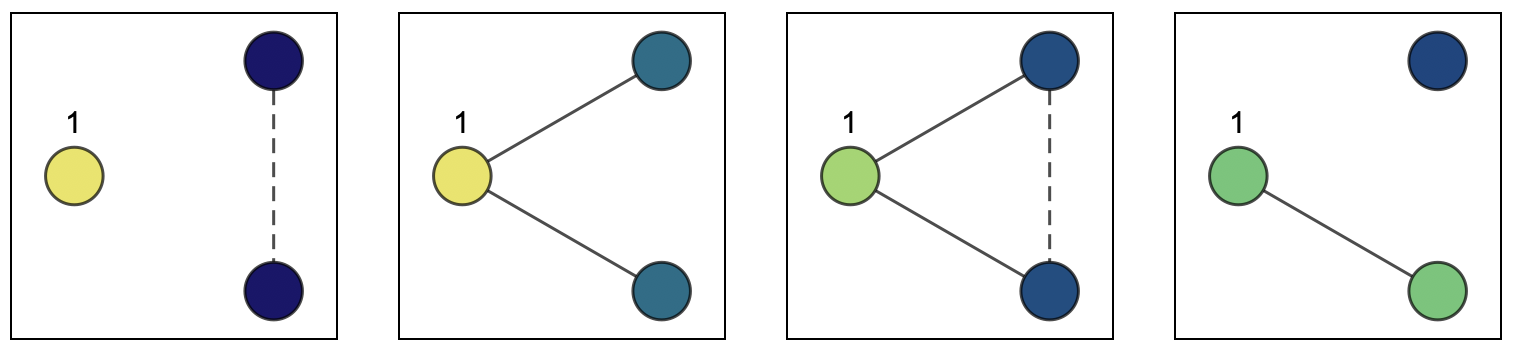}
\end{figure}
\noindent 
The first scenario is the \textit{schadenfreude} structure, in which agent 1 takes numerical pleasure in the suffering that the other agents are inflicting upon each other. Their conflict is causing agent 1's PrinceRank to increase. Some realists call this pattern bloodletting, which is when a state happily watches two rivals weaken each other. We've seen the second hub-and-spoke structure before; it's every agent's fantasy. In the third scenario, agent 1 has an ally, resulting in growth. The fourth is a composite of the first three: agent 1 has allies who are helping it grow, but the allies are weakening each other. 

Now let's consider the triadic structures that tend to be among the least preferred, using the same  parameters.
\begin{figure}[H]
\centering
\includegraphics[scale=.5]{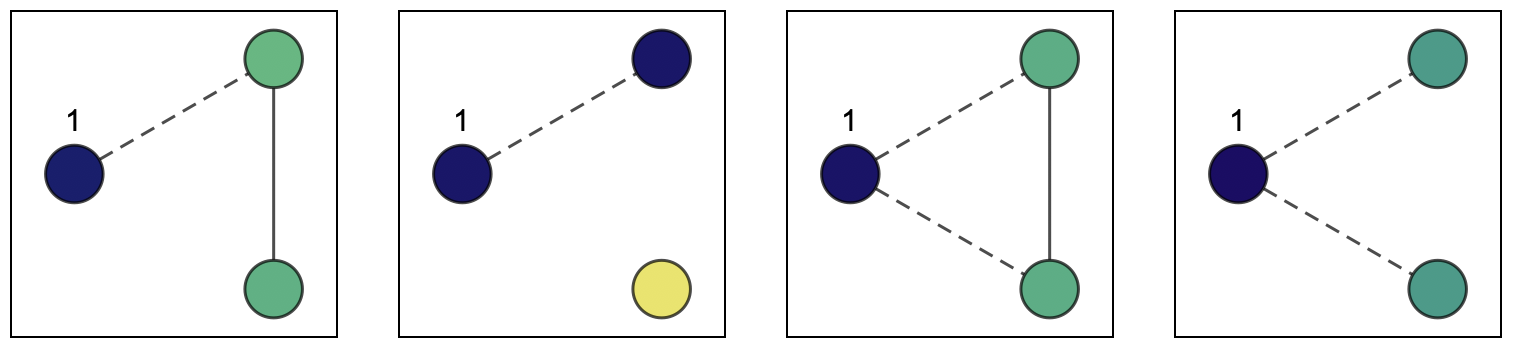}
\end{figure}
\noindent 
What these power structures have in common is that agent 1 is being attacked, sometimes when the other agents have formed an alliance. 

These examples align fairly well with intuition. In the most preferred scenarios, agent 1 is not being attacked, there are no rivals in a position to surpass it, and it is growing in absolute terms. In the least preferred structures, agent 1 is the object of violence, including by hostile alliances.

\subsection{Dynamics in Unipolar Power Structures}

\subsubsection{Introduction}

In this section, we consider the dynamics that arise in unipolar power structures. First, though, we clarify our use of the terms unipolar and bipolar. In the international relations literature, these terms are typically defined in relation to the distribution of power (see Mearsheimer 2014). A unipolar system is thought of as one with a single state that is significantly more powerful than the others; a bipolar system is one with two states that are more powerful than the rest. 

Polarity should also take into account the pattern of alliances within the system. For instance, in the two power structures below, the distribution of power in each is identical:
\begin{figure}[H]
\centering
\includegraphics[scale=.5]{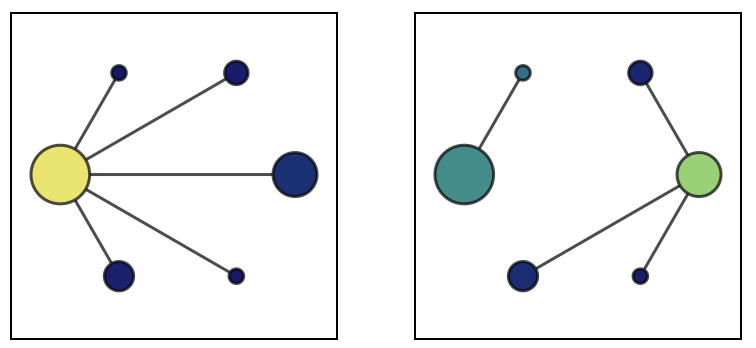}
\end{figure}
\noindent 
However, the power structure on the left is clearly unipolar (with agent 1 as the pole), and the one on the right is bipolar (with agents 1 and 4 as the poles). Below, we assume that each pole in a power structure is either a single large agent, a cluster of positively connected agents, or a substructure vaguely resembling a hierarchy (i.e. a hub-and-spoke pattern). We use terms like pole, hub, hegemon, and great power to refer to the central agents.

\subsubsection{Hierarchy Formation}

Historically, weak nations often paid tribute to a strong one, which in turn protected the weak from attack. This type of power relationship has been known to exist since the dawn of recorded history and even though it doesn't exist in the same formal sense in modern international relations, it is still the case that less powerful political entities often provide a disproportionate volume of benefits to more powerful ones, who seek to prevent the disruption of that flow of benefits. These tributary links form the backbone of hierarchical power structures.

Why do tributary relations form in the first place? When one agent is significantly more powerful than another, and the flow of mutual power balanced such that the stronger agent gets an even or disproportionate benefit out of the relationship, the stronger agent need not be threatened by the growth of the weaker one. From the perspective of the weaker agent, when resistance is futile, it's better to pay tribute that to be unaffiliated and unprotected. Under such circumstances, a cooperative tributary relationship is preferable to both parties.

When the payer of tribute is attacked by a third party, the powerful state has to decide whether to defend it. This is the dilemma faced by agent 1 below. In the diagram on the left, agent 1 does nothing, whereas on the right, it defends its tributary subject, agent 2. Considering the choice between doing nothing and engaging in conflict with the attacking party, there are size combinations in which defending one's ally is the preferred course of action:
\begin{figure}[H]
\centering
\includegraphics[scale=.5]{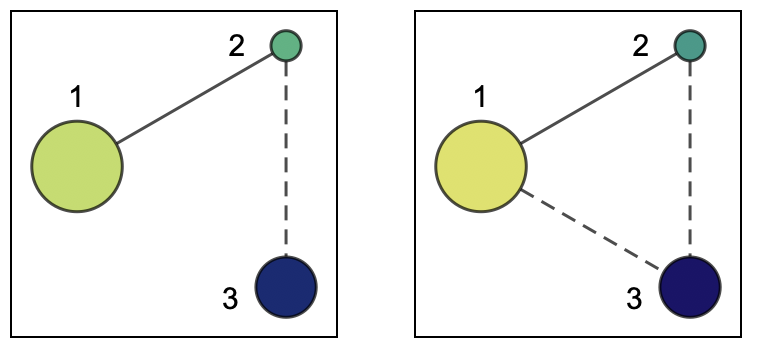}
\end{figure}
\noindent 
Above, the attacker is undermining the source of the larger agent's power. Were agents 1 and 2 to be more closely matched in size, agent 1 would have less of an incentive to defend agent 2 and might prefer to allow it to be weakened by agent 3's attack:
\begin{figure}[H]
\centering
\includegraphics[scale=.5]{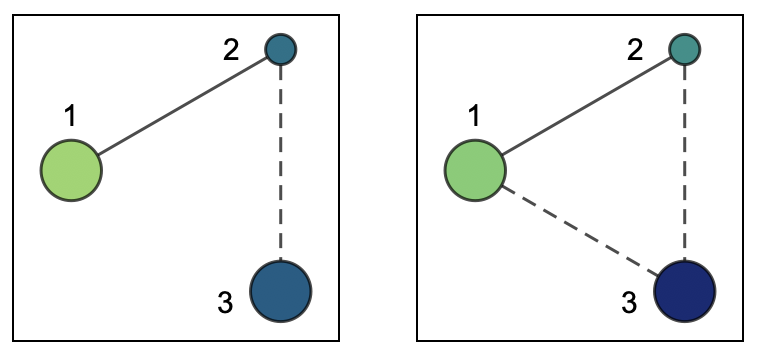}
\end{figure}

Not only is there an incentive for a powerful state to protect its tributaries, there is an incentive for tributary states to join in the powerful state's campaigns against third parties. For example, in the sequence below, agents 1 and 2 are aligned but they have conflicting policies towards agent 3 (below left). Agent 2 is faced with a choice: continue its alliance with agent 3 and lose support from agent 1 (below center), or harmonize its foreign policy with agent 1 (below right).
\begin{figure}[H]
\centering
\includegraphics[scale=.5]{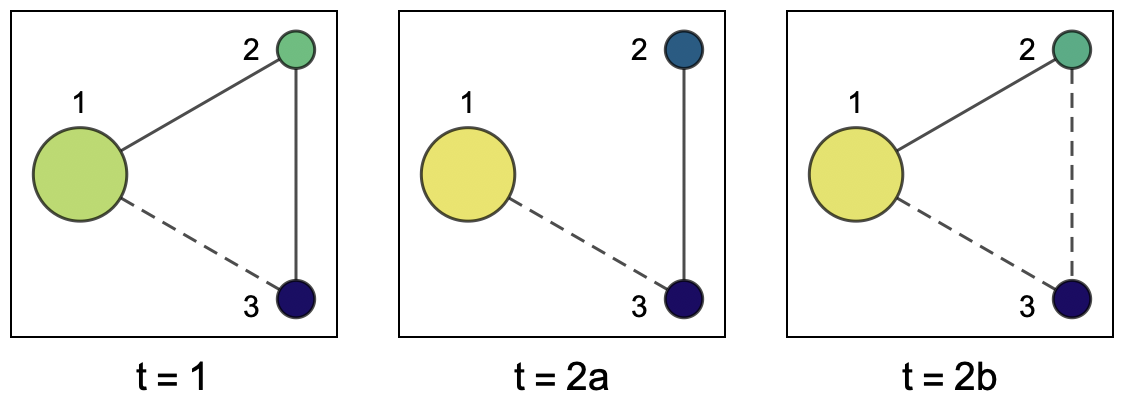}
\end{figure}
\noindent 
Agent 1's threat of time branch 2a coerces agent 2 into time branch 2b. Consequently, tributary relations create a kind of mutual defense arrangement in which a strong state defends a weak one to protect its supply of power, and the weak acts as a mercenary to the strong.

Powerful states prefer to be the exclusive recipient of tribute and will take measures to enforce that preference. Below, agent 2 is a tributary of agent 1 (left) and agents 1 and 3 (right):
\begin{figure}[H]
\centering
\includegraphics[scale=.5]{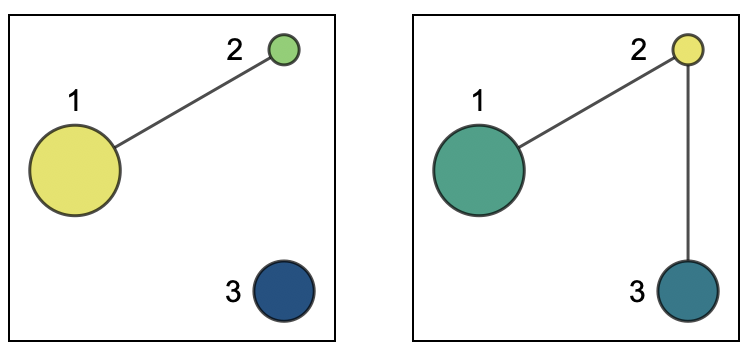}
\end{figure}
\noindent 
As shown above, agent 1 prefers to be the sole recipient of tribute.

Because tributary relations are mutually beneficial, mutually defended, and preferred to be exclusive, hierarchical power structures tend to form and persist. For other theories of network formation, see Bramoullé 2016; Easley 2010; Jackson 2008. For other approaches to state formation, see Cederman 1997.

\subsubsection{Maintaining Order}

Since at least Machiavelli, one strand of realist thought has concerned itself with the maintenance of hierarchy. The idea is that a hegemonic agent at the epicenter of a hierarchy must be vigilant about maintaining its position against possible rivals, and it must discipline coalition members in order to maintain supremacy. Given a hub-and-spoke structure dominated by a powerful agent (below left), two general challenges present themselves to the leader: coalition members fighting with each other (below center) and coalition members colluding with each other (below right):
\begin{figure}[H]
\centering
\includegraphics[scale=.5]{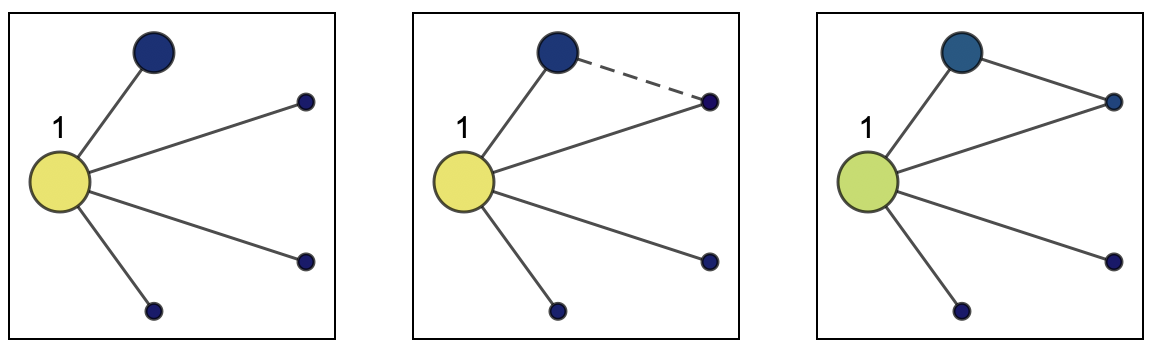}
\end{figure}

Infighting is a problem because it drains the system of constructive power that would otherwise enrich the hegemon, and collusion is a problem because it threatens the hegemon with the rise of an alternate power center. The hegemon must apply carrots and sticks to correct these situations. Temporarily withholding support from the offending parties can send a message of displeasure. In the case of infighting, sometimes it's in the hegemon's interest to threaten the aggressor with exclusion, and for the aggressor to relent (not shown). On the other hand, a hegemon may not mind infighting if it serves to weaken potential rivals and doesn't detract from the amount of power that those agents allocate to the hegemon:
\begin{figure}[H]
\centering
\includegraphics[scale=.5]{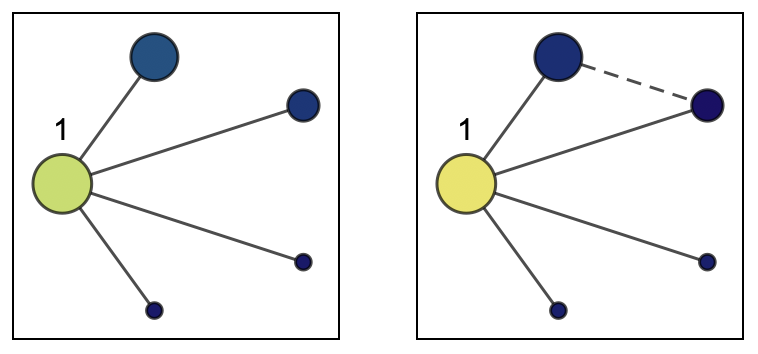}
\end{figure}

When alliance members collude, they reduce the hegemon's relative monopoly on power. Sometimes this has to be tolerated, as below, where ostracism merely causes agent 1 to do worse and the colluding agents to do better:
\begin{figure}[H]
\centering
\includegraphics[scale=.5]{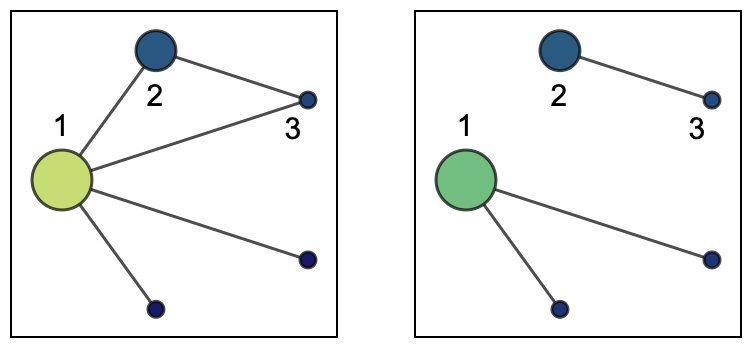}
\end{figure}
\noindent 
However, there are situations in which threatening the colluding agents with exclusion is the preferred option for the hegemon:
\begin{figure}[H]
\centering
\includegraphics[scale=.5]{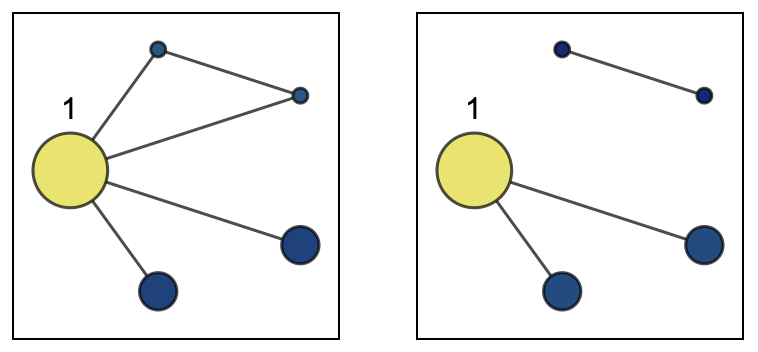}
\end{figure}
\noindent 
Above, the hegemon does better in the figure on the right (even though the colors look the same). Note that agent 1 allocates the same amount of power to agents 4 and 5 in both the left and right diagrams above.

There are situations in which more aggressive discipline is called for, here by agent 1 to deal with infighting by agent 2:
\begin{figure}[H]
\centering
\includegraphics[scale=.5]{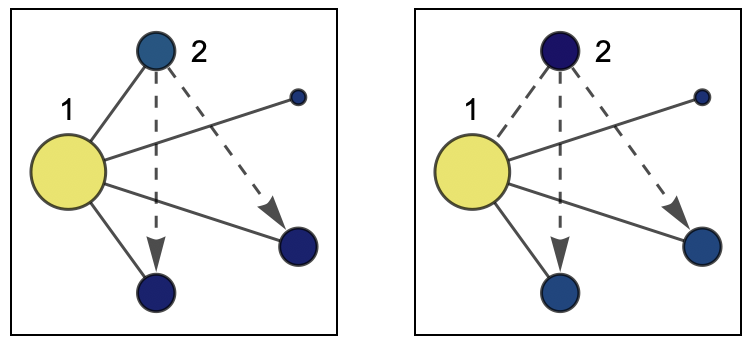}
\end{figure}
\noindent 
As shown, quantitative realism produces results that coincide with intuitive notions of hegemony maintenance.

\subsubsection{Rebellion}

The desire to rebel against a power structure, and the converse desire of powerful agents to resist disturbance, forms a theme that permeates all of history. The representation of those desires in the context of quantitative realism is straightforward. The prototypical starting point is a hub dominated by a powerful agent:
\begin{figure}[H]
\centering
\includegraphics[scale=.5]{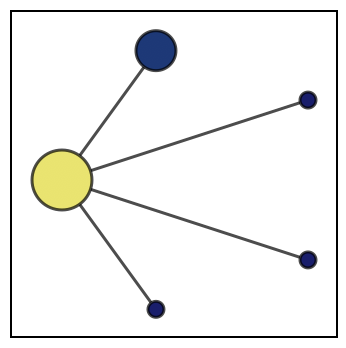}
\end{figure}
\noindent 

Rebellions are only successful when a critical mass of agents act in concert to overthrow the powerful. The figures below show how the preferences of the agents change as more parties participate.
\begin{figure}[H]
\centering
\includegraphics[scale=.5]{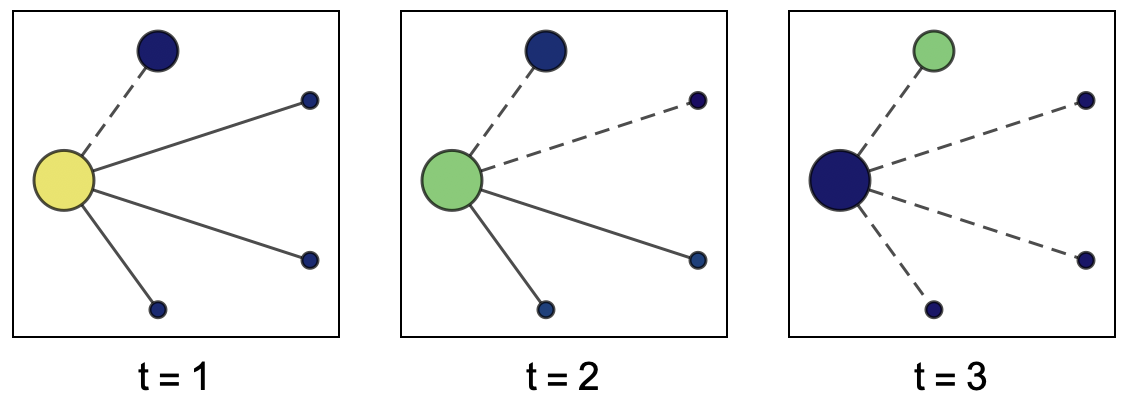}
\end{figure}
\noindent 
The PrinceRank of the hub agent plummets as more agents join the fray. The model also reflects the preference of agents to free ride on other's aggression. For example, the PrinceRank of those who don't rebel (agents 4 and 5, above left and center) is higher than it would be if no one had rebelled at all. They reap the benefits of the fight without incurring any of its costs. Moreover, the utility of those who do rebel is generally lower than it would have been had they not rebelled at all, indicating that there's a cost to rebelling. Part of the calculus of revolution is to determine whether those short-term costs are worth taking.

Additionally, the PrinceRank of a rebel goes up when more agents rebel along with it, because the hegemon then has to fight on multiple fronts and weakens more quickly. In the examples above, the largest rival, agent 2, benefits the most when everyone rebels. Not only is its PrinceRank higher than all of the other agents (above, far right), including the (soon-to-be former) hegemon, it is also higher than it was in the original, hierarchical power structure. Agent 2 is the only one who would prefer a full rebellion, possibly suggesting a more general principle that beta agents tend to have the strongest incentive to overthrow the regime.

Rebels have an incentive to cooperate with each other. Mutual support can be advantageous to them, but the hegemon has incentives to maintain order (Section 4.3.2). For example, even though it's hard to discern below, the smaller agents have incentives to support agent 2 as a counterbalance to 1.
\begin{figure}[H]
\centering
\includegraphics[scale=.5]{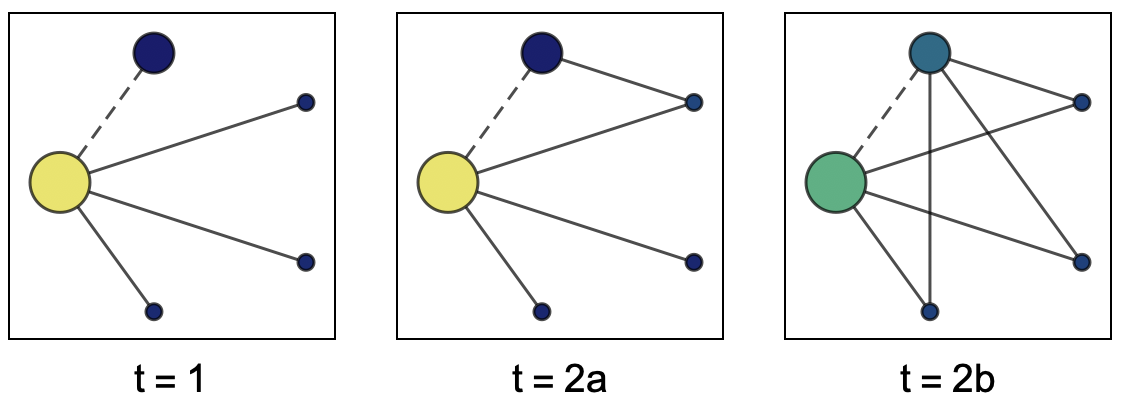}
\end{figure}
 
These arrangements are likely to be short-lived if agent 1 has the ability to quell the uprising and reimpose order. It seems intuitive that a threatened hegemon would want to fight back. However, it could also increase the power allocated to the disgruntled agents, essentially buying their loyalty. (These scenarios are not depicted.)

The final stage of a successful rebellion or revolution is when a new political order is established. Suppose we play out the initial scenario, with all of the agents rebelling simultaneously (below left):
\begin{figure}[H]
\centering
\includegraphics[scale=.5]{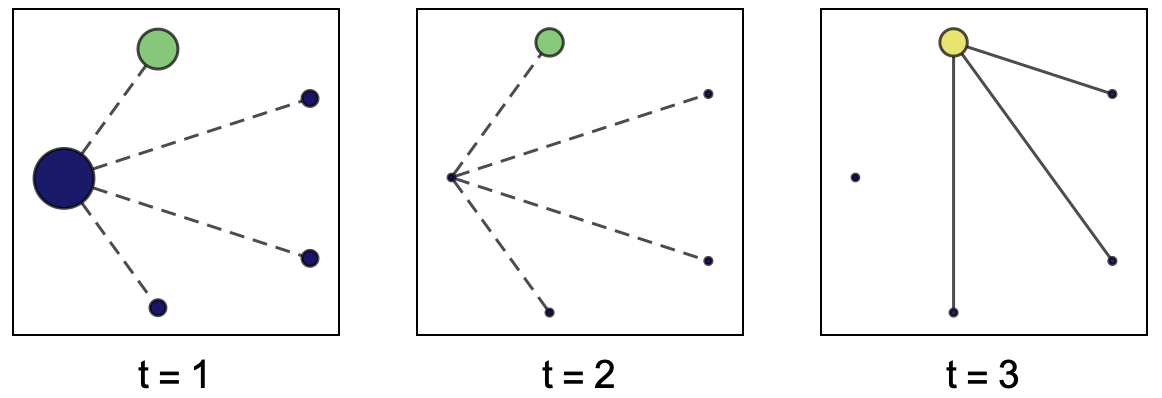}
\end{figure}
\noindent 
After the hegemon is defeated (above center), the victors form a new hierarchy (above right) out of the power vacuum (see Section 4.5). In this particular example, all of the rebels end up with a higher PrinceRank than they did when the revolution began, despite the fact they they have all been diminished in absolute size. This is because they have eliminated agent 1, a giant who was dominating them.

Quantitative realism can account for the basic dynamics of revolution: agents willing to absorb the short-term pain of fighting in order to achieve the long-term benefits of a newly balanced order, and established powers that struggle to maintain the status quo.

\subsubsection{Disintegration and Collapse}

The lifecycle of a unipolar power structure eventually ends in disintegration and collapse. Broadly speaking, there are two types of triggers: exogenous events that shock the system from the outside, and endogenous ones that arise from altered incentives within an evolving network. In both, the withdrawal of support for the central node creates centrifugal tendencies that pull the network apart.

A pure hierarchy is stable. The hub tends to grow at a faster rate than the tributary states and can dole out just enough power to each tributary to keep it in the network, while keeping other power in reserve for the maintenance of order. However, most real world networks are not pure hierarchies. Smaller states in a unipolar structure tend to have relationships with each other, both positive and negative, and the leader may not be fully connected to all of the smaller agents (below left). Suppose there is an exogenous change to such a power structure. For example, a central agent might weaken due to epidemic, natural disaster, military attack, or economic crisis (below right).
\begin{figure}[H]
\centering
\includegraphics[scale=.5]{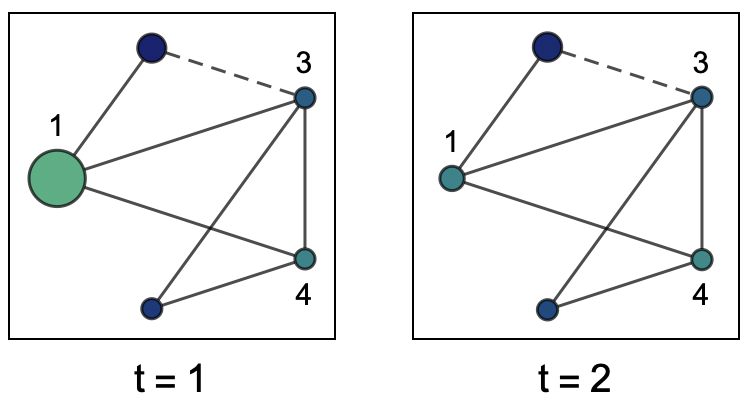}
\end{figure}
\noindent 
In the sequence above, agent 1, initially powerful (left), has weakened due to some outside event. The system at t=2 (above right) has now become a messy, vaguely bipolar structure, with agents 1 and 4 positioned as rival poles. Agent 3 is an ally to both poles, a situation not likely to persist (see Section 4.3.2). After agent 3 chooses a side (below left), agents 4 and 5 are then better off withdrawing into a separate alliance (below right).
\begin{figure}[H]
\centering
\includegraphics[scale=.5]{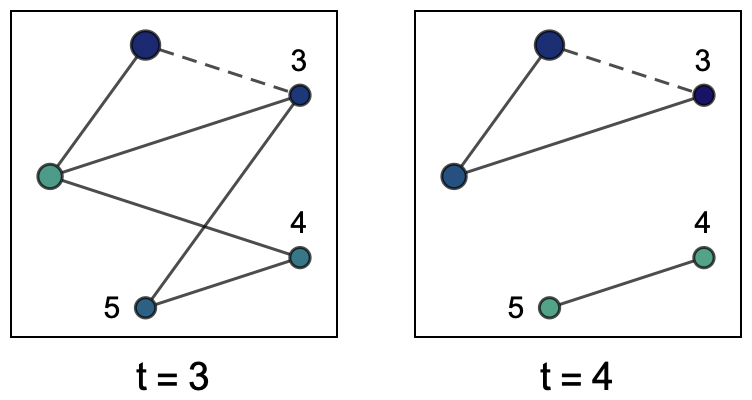}
\end{figure}
\noindent 
This sequence indicates how unipolar structures can decompose when the hub agent is suddenly weakened. It's obviously a caricature, as it doesn't take into account all of the possible agent decisions, only those of the more influential agents. This sketch should, however, convey the gist of the disintegration process. A similar process (not illustrated) can happen when a tributary state suddenly becomes more powerful, such as through the discovery of natural resources or the development of a new technology.

Disintegration can also be triggered by the uneven growth of agents in the network. For instance, in the network below (left), agent 1 is dominant in size and relatively central, but agent 4 is well-connected, making its small size deceptive. If these relationships were to remain constant, over time the law of motion will cause agent 4's size and PrinceRank to increase disproportionately (below right).
\begin{figure}[H]
\centering
\includegraphics[scale=.5]{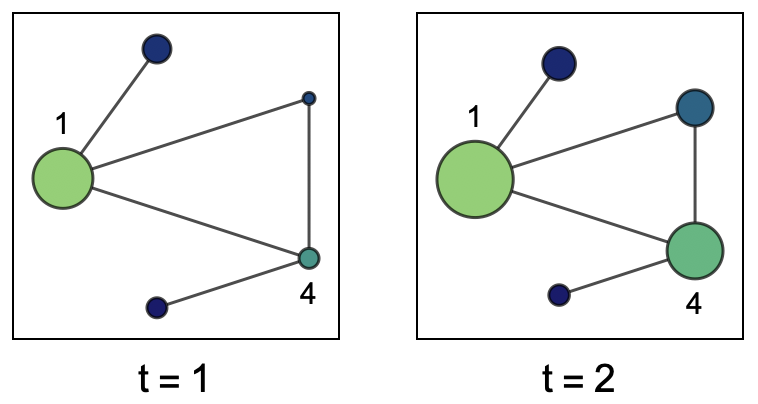}
\end{figure}
\noindent 
At t=2, agent 4 has become a fearsome rival to agent 1, creating a system with multiple poles that is likely to disintegrate along the lines of the reasoning in the example above.

\subsection{Dynamics in Bipolar Power Structures}

\subsubsection{Introduction}

In this section, we consider bipolar power structures, for example:
\begin{figure}[H]
\centering
\includegraphics[scale=.5]{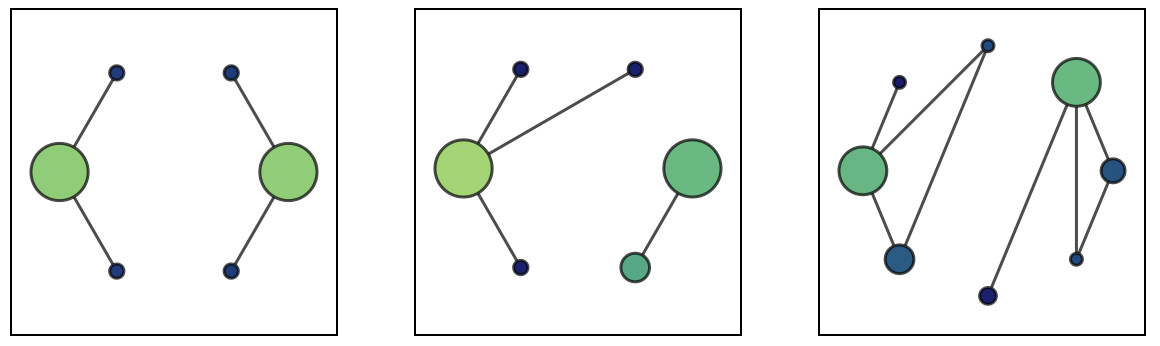}
\end{figure}
\noindent 
As discussed in Section 4.3.1, a bipolar power structure is a system characterized by two clusters or hierarchies of agents, each of which has a relatively powerful agent at its epicenter.

\subsubsection{Tributary Realignment}

Weaker members of a bipolar system face incentives as to which cluster of agents to be a part of. For the reasons discussed in Section 4.3.2, alpha agents do not like when their tributary states serve two masters. So weak agents must usually choose to belong to one side or the other of a bipolar world.

In the scenarios below, agent 3 can choose between joining agent 1's alliance (below left) or agent 4's (below right).
\begin{figure}[H]
\centering
\includegraphics[scale=.5]{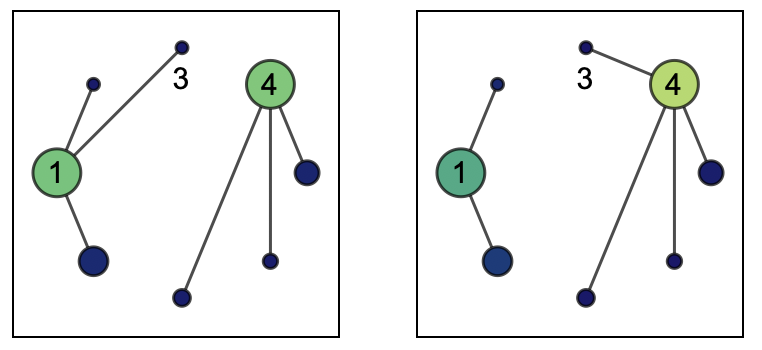}
\end{figure}
\noindent 
Each hegemon would prefer that agent 3 join its gang. For them, every alternate alignment is a double loss: less power for themselves, and more power for a competitor. Here it would be better for it to align with agent 1. However, the choice might be different if agent 4 was more powerful than agent 1.

What happens if there's an incentive for the tributary state to defect? For example, going back to the original dilemma that we created for agent 3, what options does agent 4 have if agent 3 shifts allegiance to agent 1? One option is for agent 4 to redistribute more of its power toward agent 3 and away from its other client states, thus making it more lucrative for agent 3 to remain in the alliance. Such a redistribution would be disliked by agents 5, 6, and 7, and could alter their own deliberations about which hegemon to align with.

If carrots don't work, there's always the stick. As the diagrams below illustrate, the losing hegemon (agent 4) may have an incentive to attack the defector (agent 3):
\begin{figure}[H]
\centering
\includegraphics[scale=.5]{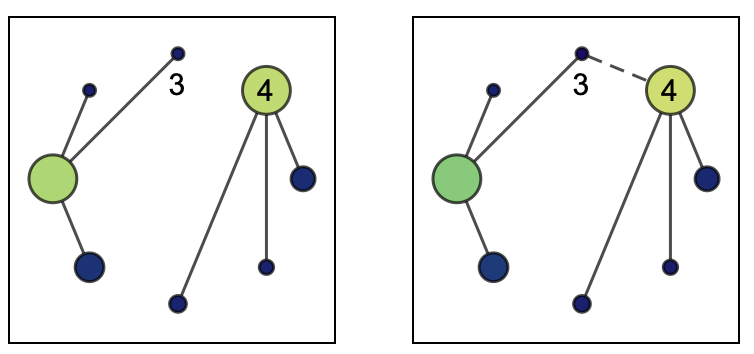}
\end{figure}
\noindent 
In the figure above, agent 4's PrinceRank goes up slightly when it attacks agent 3. To the extent that the possibility for this sort of retaliation is known to all parties, it may keep the tributary states in line.

\subsubsection{Divide and Rule}

Divide and rule (divide and conquer) is a classic strategy of political control. It entails breaking up larger concentrations of power by sparking rivalries and preventing smaller agents from linking up. In the model presented here, an agent can't interfere directly with the relationship of two other parties, and it can't create dissension by spreading lies. However, agents can nonetheless choose tactics that use rewards and punishments to break up rival coalitions. 

What kind of opposing coalition would an agent would prefer to deal with? As the figure below indicates, agent 1 would generally prefer a group of rivals to be divided by infighting (right) rather than unified against it (left), all else being equal.
\begin{figure}[H]
\centering
\includegraphics[scale=.5]{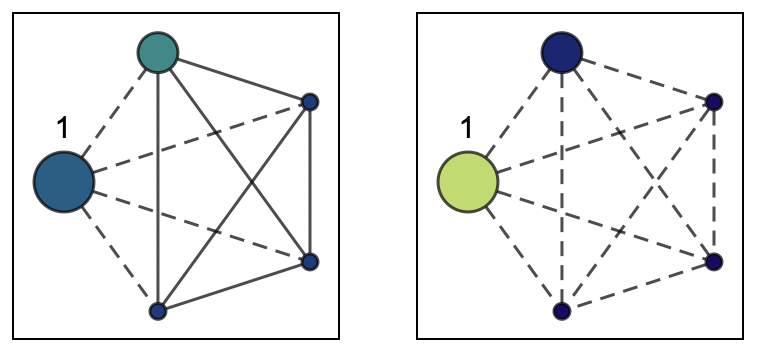}
\end{figure}
\noindent 
Assuming that an agent finds itself in a situation like the one on the left, how might it attempt to turn it into one more like that on the right? One possibility is to introduce discord into the coalition by attacking some members and co-opting others. This strategy can imbalance relationships, triggering waves of strife and punishment within the coalition, because the co-opted agents have an incentive to turn on their leader and other agents are forced to choose sides.

Let's examine the simplest possible case of divide and rule (below). Suppose that agents 2 and 3 have an alliance, and that agent 1 is stronger and not included (far left):
\begin{figure}[H]
\centering
\includegraphics[scale=.5]{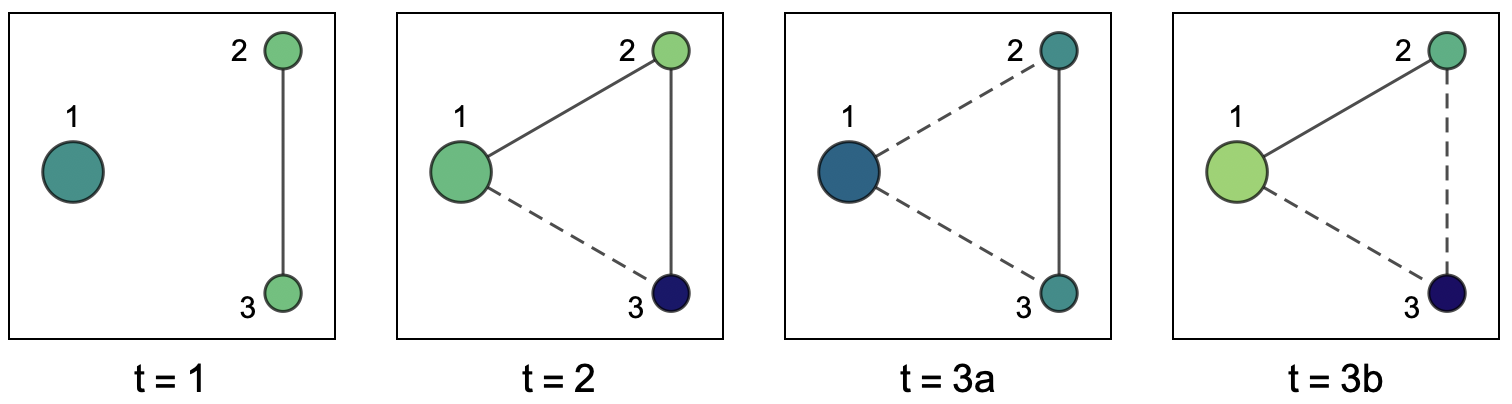}
\end{figure}
\noindent 
Agent 1 commences divide and rule operations by co-opting agent 2 and attacking agent 3 (center left). Agent 2 then has to choose between two outcomes: remain with agent 3 and be attacked by agent 2 (center right), or betray agent 2 and align with agent 1 (far right). As agent 2's PrinceRank indicates, the latter option is preferred, and agent 1 has accomplished its objective of fracturing the opposition. Agents can also use divide and rule to exploit existing divisions within an opposing coalition, causing other members to get sucked into the fray and diverting destructive power away from the instigating agent.

Even though we've glossed over some of the finer points (for example, the sizes of the agents here do not change at the various time steps), it should be evident that quantitative realism illuminates the essential logic underlying divide and rule: both the preference structure that motivates it and the mechanism of its attainment.

\subsubsection{Latent Tension}

Some power structures have intrinsic stress due to the incentives of certain agents to initiate conflict. We explore this latent tension by looking at power structures in which there is no active conflict, but where a party can increase its PrinceRank by starting one. We'll focus on bipolar power structures and examine the tensions between the two great powers. Consider the following power structure, in which the great powers that have a slight size difference. Agents 1 and 4 can form one of the three possible relationships:
\begin{figure}[H]
\centering
\includegraphics[scale=.5]{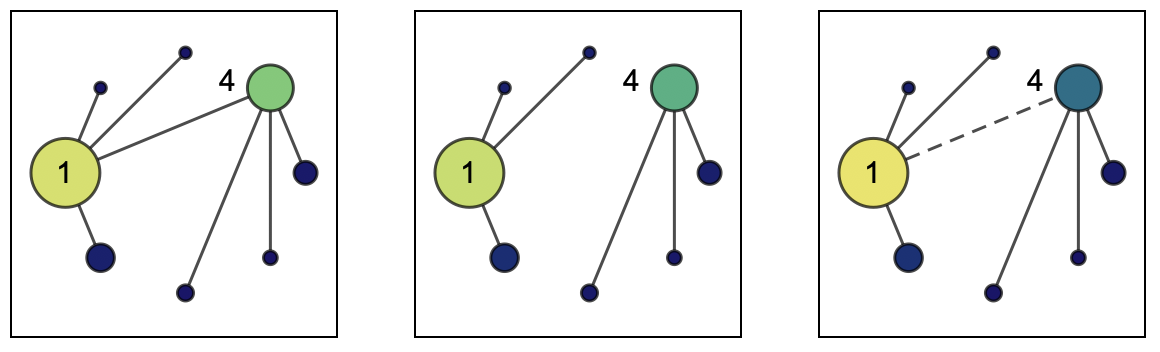}
\end{figure}
\noindent 
The structure above (center) has a latent tendency for great power conflict because agent 1 can improve its PrinceRank by instigating a fight with agent 4 (above right). It doesn't matter that agent 4's PrinceRank is lowest in this scenario, because we assume that if one agent wants to fight, a fight will occur. Accordingly, there is an instability inherent within this power structure that is not immediately obvious on the surface.

Compare the situation above with a variation in which the two most powerful agents are roughly equal in size (below center):
\begin{figure}[H]
\centering
\includegraphics[scale=.5]{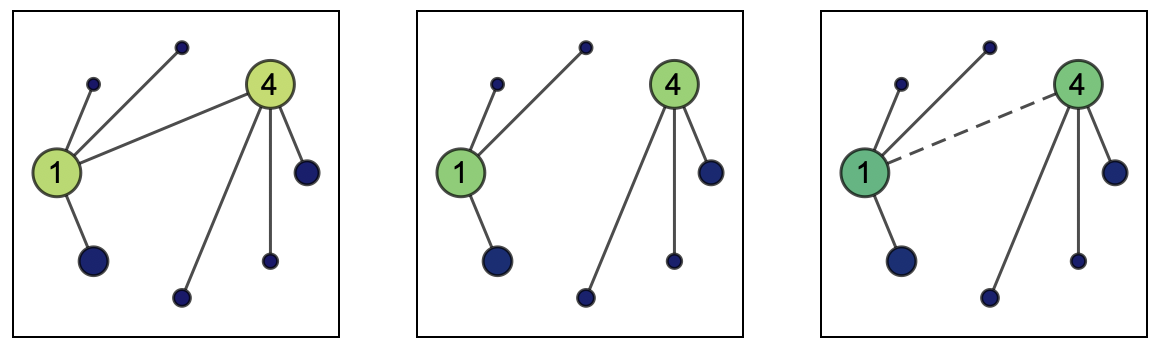}
\end{figure}
\noindent 
In the diagrams above, agent 1 and its allies are about as powerful as agent 4 and its allies, and neither of the great powers has an incentive to fight with each other. In fact, it's in both of their interests to cooperate with each other (above left). So there is less latent tension for conflict in the scenario above center (other agents may have incentives to fight, but the great powers do not).

Obviously, the same kind of tension can exist between any two agents in a system, and not just between great powers. It doesn't mean that two agents will necessarily fight merely because some structural tension is present. Agents also have to consider all of their other possible relationships and any counter-reactions to their actions. So the depiction above is not meant to suggest that conflict is inevitable under any given set of circumstances. However, power structures with latent tension do illustrate the basis of the fear and insecurity that agents experience in an anarchic setting. Quantitative realism accounts for the fact that power structures that appear to be in equilibrium may actually be fraught with destructive potential.

\subsubsection{Conquest}

Conquest in the real world ordinarily entails the victor acquiring the resources of the vanquished. Here, resources are not a factor in the calculations of the agents. However, we can nonetheless address this topic by treating tributary states as if they are booty to be won in the aftermath of a conflict. We'll take as a starting point a bipolar power structure which, as discussed in Section 4.4.4, has a latent tendency for conflict between agents 1 and 4:
\begin{figure}[H]
\centering
\includegraphics[scale=.5]{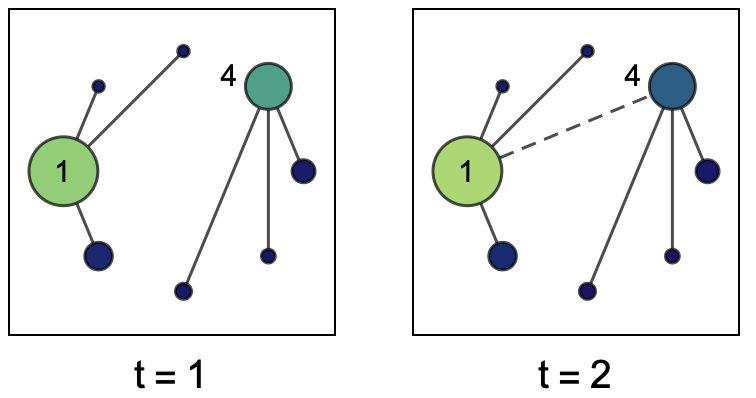}
\end{figure}
\noindent 
Suppose that this conflict occurs, with some of agent 1's proxies joining in. They don't have a direct incentive to do so, in terms of PrinceRank, but they could be coerced into it if agent 1 were to threaten to withdraw support for them. 
\begin{figure}[H]
\centering
\includegraphics[scale=.5]{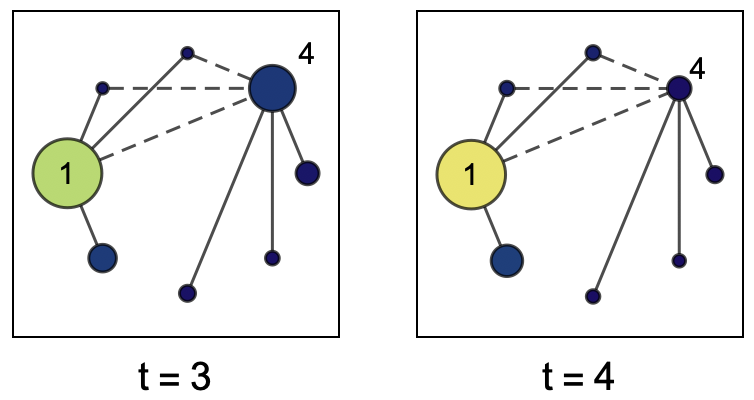}
\end{figure}
\noindent 
As the battle unfolds, agent 4 is diminished (above right and below left). At this point (t=5), agent 4 is better off submitting and paying tribute to agent 1 (below right), and agent 1 happy to accept it, assuming that agent 4's former tributaries also switch their allegiance to agent 1. Doing so is in their interest as well:
\begin{figure}[H]
\centering
\includegraphics[scale=.5]{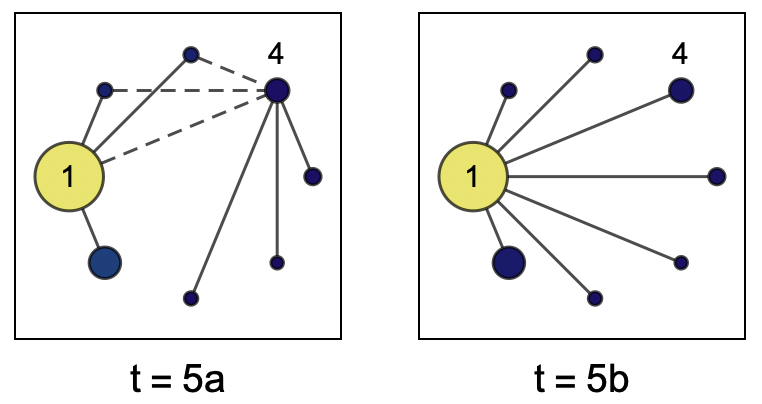}
\end{figure}
\noindent 
This is a stylized example of conquest, the object of which is to eliminate a rival and obtain their material, represented here by subservient agents with little freedom of action.

\subsubsection{Balancing, Bandwagoning, and Buck-Passing}

Next we examine balancing, bandwagoning, and buck-passing, which can be represented in power structures. Consider a scenario in which there is one hierarchy, led by agent 2, and a separate, rival structure:
\begin{figure}[H]
\centering
\includegraphics[scale=.5]{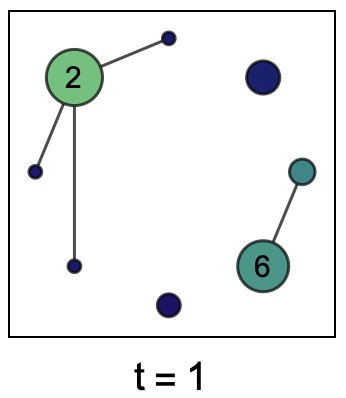}
\end{figure}

Balancing, bandwagoning, and buck-passing are various ways that an agent can respond to rivals. Let's look at the three behaviors from the perspective of agent 2 to see how it might react to its primary rival, agent 6. Balancing can be construed in two different ways: as the initiation of aggression to cut down a rival (below left), or as the formation of an alliance that counterbalances the rival by building strength (below right):
\begin{figure}[H]
\centering
\includegraphics[scale=.5]{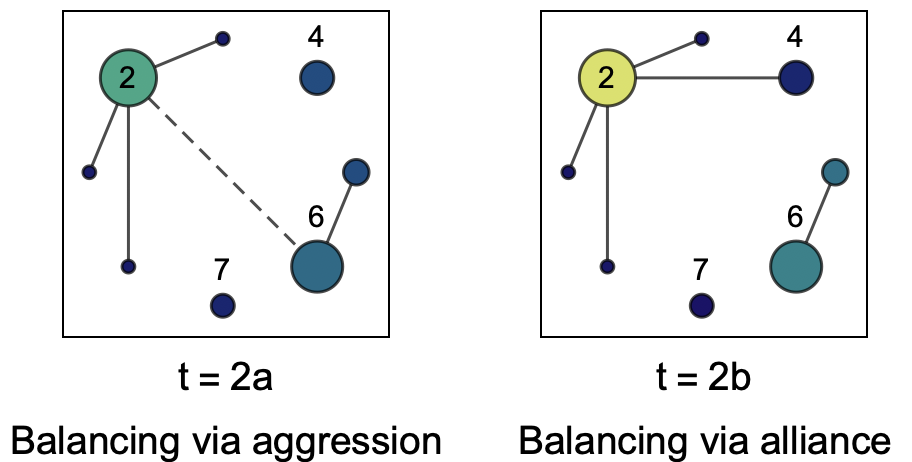}
\end{figure}
\noindent
Bandwagoning, in contrast, is when an agent chooses to cooperate with a rival (below left). And in buck-passing, the agent prefers that a third party do the dirty work of balancing against the rival (below right).
\begin{figure}[H]
\centering
\includegraphics[scale=.5]{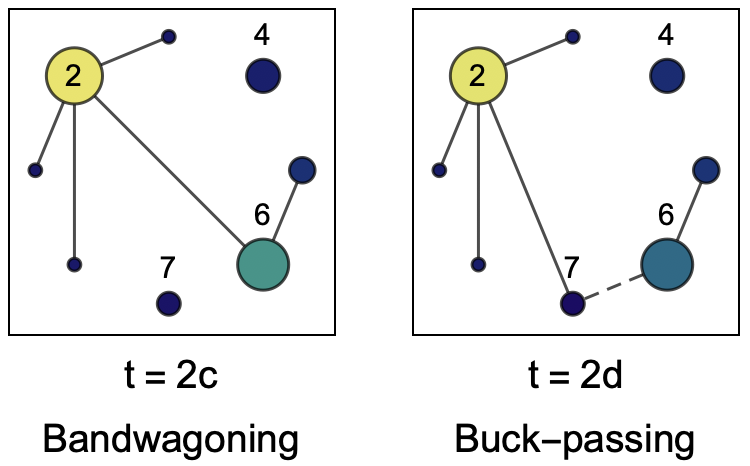}
\end{figure}
\noindent
Of the options presented for the scenario above, which is the best choice for agent 2? Balancing via aggression would lower agent 2's PrinceRank compared to the baseline case (t=1), so that's not a preferred choice. Bandwagoning lowers agent 6's PrinceRank, suggesting that it would probably not agree to cooperate with agent 2 under the circumstances. Agent 7 doesn't make out very well as a conduit for buck-passing, and is unlikely to agree to such a mercenary role unless agent 2 allocated more power to it. Balancing via alliance seems like the strongest move for agent 2 (again, of the options shown), and agent 4 has an incentive to go along with it.

This is a simplistic case, but it shows how balancing, bandwagoning, and buck-passing can be analyzed in the context of quantitative realism. Quantitative realism offers a way to resolve longstanding questions within balance of power theory about which of these behaviors are preferred in various types of power structures.

\subsubsection{Chain-Ganging}

In this section, we explore a phenomenon known in international relations as chain-ganging. In it, an isolated bilateral conflict spirals out of control, dragging in third parties who are aligned with the original combatants. World War I is a quintessential example: the European great powers were eventually sucked into what began as an isolated dispute between Serbia and the Austro-Hungarian Empire. 

Suppose that a conflict erupts between the tributary states of two great powers:
\begin{figure}[H]
\centering
\includegraphics[scale=.5]{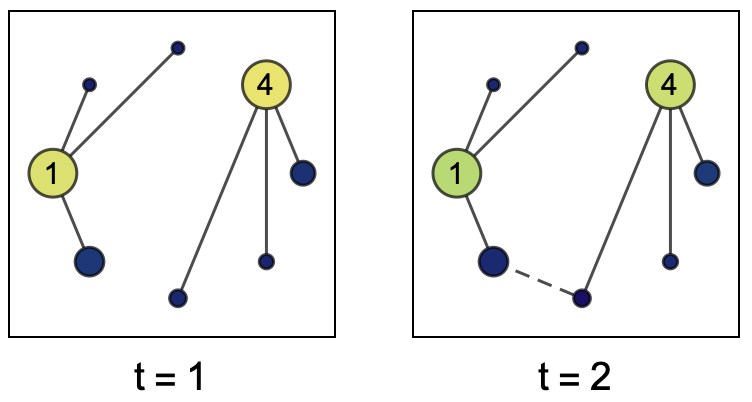}
\end{figure}
\noindent
The two hegemons would prefer that this conflict not occur (their PrinceRank was higher before the fight started), and it may be that they can rein in the aggressors by withdrawing support for them (below left):
\begin{figure}[H]
\centering
\includegraphics[scale=.5]{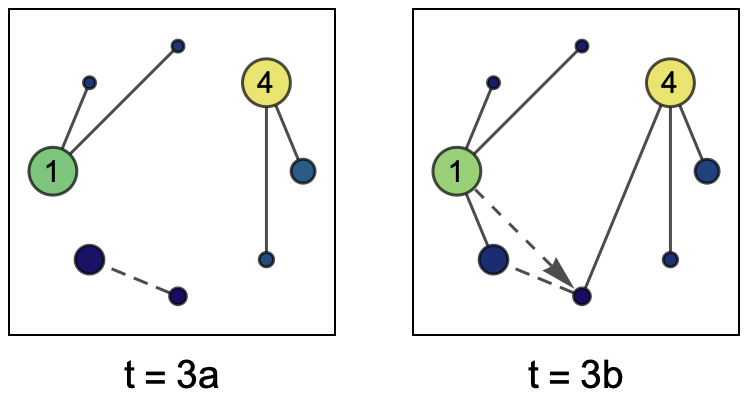}
\end{figure}
\noindent
However, as the scenario above left illustrates, it is not in agent 1's interest to do this. Nevertheless, the combatants do much worse in t=3a after they've been isolated, and it may be that the threat of such a consequence is enough to end the hostilities. As an alternative (t=3b, above right), suppose that one of the hegemons cannot discipline its client state, and instead decides to defend its ally by attacking the enemy agent. In that case, it is in agent 1's interest to engage in such a defense, and in response it starts to be in agent 4's interest to join the fray as well:
\begin{figure}[H]
\centering
\includegraphics[scale=.5]{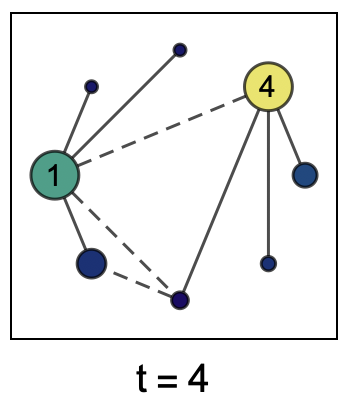}
\end{figure}
\noindent
In the diagram above, agent 4 has started a direct conflict with agent 1, and from here the conflagration is likely to become totalistic until the distribution of power is so altered that a new calculus of interests eventually arises. No one will be better off in absolute terms by the time the conflict burns out, and probably all or most parties will have lower PrinceRanks compared to before the madness began. Nonetheless, there's a logic to chain-ganging that is difficult to escape, as history has shown all too well.

\subsubsection{Unity Under Threat}

Another theme in power politics is that agents tend to unify in the face of a common threat. For example, consider the four power structures below:
\begin{figure}[H]
\centering
\includegraphics[scale=.5]{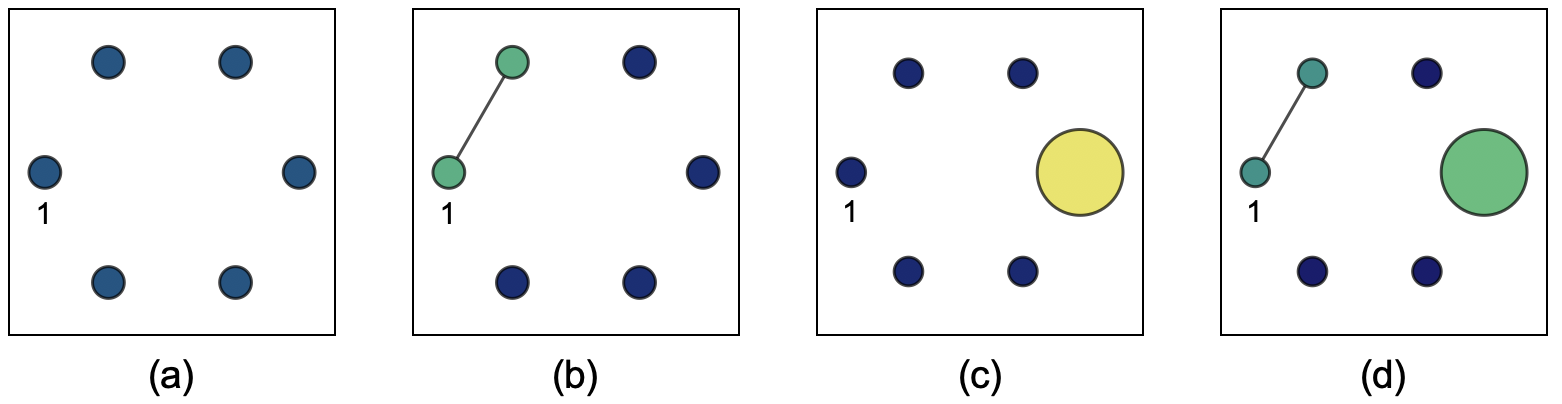}
\end{figure}
\noindent
In diagrams (a) and (b), in which all agents are the same size, when two agents unify it increases their PrinceRank by a factor of 2.7. In contrast, in diagrams (c) and (d), in which there is a more powerful agent in the system, cooperating increases PrinceRank by a factor of 4.7. So the mere presence of a dominant agent intensifies the incentives of smaller agents to unite with each other. (This sort of four-way comparison is necessary here because agents generally like cooperation and prefer to be the hub of the network, and we are trying to show that the presence of a more powerful agent has the causal effect of amplifying that desire.)

The same incentive holds true when the powerful agent is not merely present but also behaving aggressively towards weaker ones:
\begin{figure}[H]
\centering
\includegraphics[scale=.5]{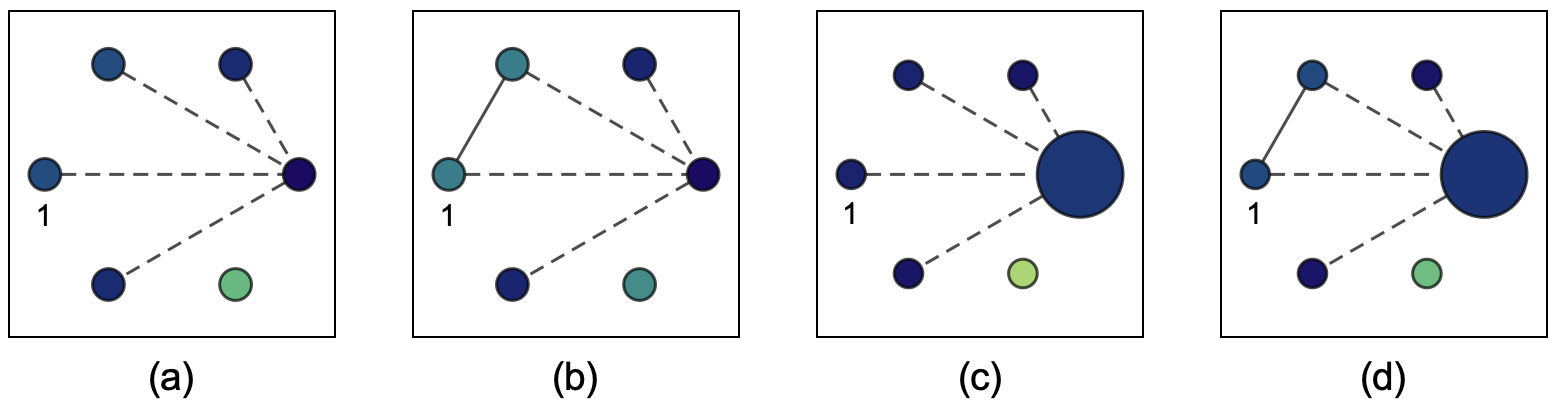}
\end{figure}
\noindent
In the scenarios above, agent 4 is now attacking agents 1 and 2. Agent 1's PrinceRank increases by a factor of 1.4 when transitioning from (a) to (b), and a factor of 1.8 when going from (c) to (d). Note that this assumes that the power that the smaller agents allocate to cooperation does not detract from the power needed to defend themselves against their attacker.

\subsection{Power Vacuums}

The term power vacuum, as used both here and in common parlance, denotes a situation that is not (or is no longer) dominated by a powerful agent.\footnote{ We distinguish power vacuum from anarchy, the former being the absence of structure and the latter being the absence of authority (e.g. institutions, laws, and customs).} It is characterized by agents who are approximately equal in size and with interrelationships that are diffuse and lacking in coherence. There may be pockets of order within a power vacuum, for example, trivial alliances or conflicts. However, from a global perspective, vacuums are more or less structureless and short-lived.

\begin{figure}[H]
\centering
\includegraphics[scale=.5]{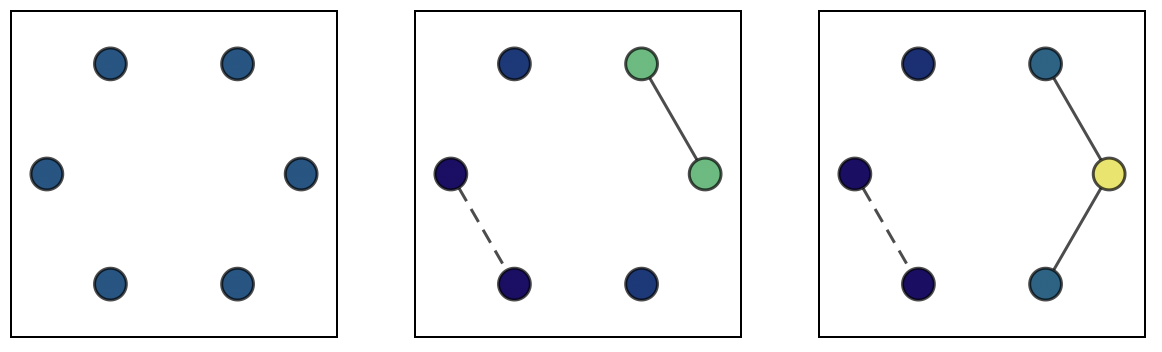}
\end{figure}
\noindent 
The structure on the left is an extreme vacuum: everyone is the same size and there are no relationships. The other two figures show increasing structure, but still little compared to the unipolar and bipolar systems considered in the sections above. 

We would not expect power vacuums to last very long, and the diagrams above suggest why. In the first place, the agents that are starting to form coalitions develop an advantage over those who have no relationships and, even more so, over those embroiled in conflicts, so there are rewards for those who associate positively with others. A power vacuum presents an opportunity for an agent to become a power center in the emerging system.

Second, in the absence of one or more great powers, there is an increased likelihood of random violence among the agents, as no one has the strength to create an atmosphere of deterrence by punishing aggressors. Larger agents have stronger incentives protect their allies than smaller agents do, and more capacity to do so (Section 4.3.2). This is perhaps the quantitative version of the principle articulated by Hobbes in \textit{Leviathan}: the need for a strong central power to prevent the ``war of all against all."

These two pressures, the competition for agents to become dominant and the preference of agents for organized domination instead of undeterred violence, make power vacuums transient and hierarchies likely to emerge from them. Quantitative realism provides a way to reason about this process.

\section{Conclusion}

We have set forth a quantitative interpretation of political realism. Starting from six postulates capable of mathematical representation, we have shown agents that exhibit realist behavior. The results are admittedly tentative and impressionistic, yet suggestive.

This formulation of realism is deliberately abstract, and could be made more lifelike in two obvious ways. First, it currently does not include any notion of geographical distance. Presumably results would better reflect historical outcomes if the model included some sort of decay factor that caused the effects of power to be attenuated as a result of being projected over long distances. Second, the power struggles here are not over any thing in particular, except power itself. In the real world, power struggles are often driven by resource scarcity. This model could be extended to allow agents to possess resources, by representing them as nodes that emit constructive power to whoever possesses them.

Even without such enhancements, this framework advances the current thinking on political realism in several ways. First, the approach is axiomatic, setting forth foundational ideas from which specific consequences can be derived. Second, it is quantitative, treating power, in both its stationary and moving states, as something measurable. Third, it considers the interrelationships among political agents, and not just their relative capacities. Fourth, it provides a dynamic account of power struggles: changing interrelationships cause power levels to rise and fall, and changing power levels cause agents to adjust their interrelationships in response. Fifth, by including a mechanism for cooperation, it helps account for the emergence of order and hierarchy.\footnote{ This may even resolve the core tension between realism and liberalism.} Sixth, it offers analytic continuity between the external behavior of nation-states and the domestic political pressures shaping their foreign policies, to the extent that both fora involve an integrated network of political actors. Finally, it is potentially generalizable beyond the context of nation-states.\footnote{ At a minimum, the proposed framework generalizes realism beyond the great powers, which has been its traditional object of study (Mearsheimer 2014).}

Realism predicts what will happen when certain preconditions are met, specifically, when agents with the capacity for violence and mutual assistance interact under anarchy. It is unlikely that there has ever been a time in human history when those conditions have been met perfectly, as cultural norms, institutions, and the like have arguably always tempered and channeled the raw struggle for power. Political realism means that, to the extent that those preconditions do exist, certain vaguely predictable processes are likely to emerge. As such, there is no reason that political realism should be limited to its traditional sphere of international relations.\footnote{ "Realism merely requires anarchy; it does not matter what kind of political units make up the system. They could be states, city-states, cults, empires, tribes, gangs, feudal principalities, or whatever." (Mearsheimer 2001)} It describes a process that will tend to occur whenever conditions allow, regardless of whether the context is an ancient empire, the dissolution of a government into anarchy, a regime of slavery, a slum, a street gang, an open-ended survival game (Groen 2015), or interplanetary conflict. Because the theory's starting assumptions are expressed generally, we should expect the dynamic behavioral processes that follow from them to be generally applicable.

This model is less an assertion about how the world is than it is a description of what realism could be. It describes the subject of realism: power. It quantifies the essential variables of realism: the capacities of states and their interrelationships. It describes the effect of those variables: the flow of power. It has the potential to answer the questions that realism asks: how agents use their power in the context of struggles over it. And it provides a single, unifying framework for a variety of phenomena of traditional concern to realism. As such, it provides a methodology for resolving longstanding debates by rendering the subject quantitative.

\theendnotes

\section{References}

\begin{footnotesize}
\begin{hangparas}{.2in}{1}

Axelrod, Robert. 1995. ``A Model of the Emergence of New Political Actors" in Nigel Gilbert and Rosaria Conte (eds.). \textit{Artificial Societies: the Computer Simulation of
Social Life}. London: University College Press.

Axelrod, Robert and D. Scott Bennett. 1993. ``A Landscape Theory of Aggregation." \textit{British Journal of Political Science}, Vol. 23, No. 2, pp. 211-233. 

Axelrod, Robert. 1984. \textit{The Evolution of Cooperation}. Basic Books.

Bramoullé, Yann, Andrea Galeotti, and Brian Rogers. 2016.  \textit{The Oxford Handbook of the Economics of Networks}. Oxford University Press.

Bueno de Mesquita, Bruce, et al. 2004. \textit{The Logic of Political Survival}. MIT Press.

Buzan, Barry and Richard Little. 2010. ``World History and the Development of Non-Western International Relations Theory" in Acharya, Amitav and Barry Buzan. 2010. \textit{Non-Western International Relations Theory: Perspectives On and Beyond Asia}. Routledge.

Cederman, Lars-Erik. 1997. \textit{Emergent Actors in World Politics: How States and
Nations Develop and Dissolve}. Princeton University Press.

Easley, David and Kleinberg, John. 2010. \textit{Networks, Crowds, and Markets: Reasoning about a Highly Connected World}. Cambridge University Press.

Glaser, Charles L. 2010. \textit{Rational Theory of International Politics: The Logic of Competition and Cooperation}. Princeton University Press.

Goddard, Stacie E., and Daniel H. Nexon. 2016. ``The Dynamics of Global Power Politics: A Framework for Analysis." \textit{Journal of Global Security Studies}, 1(1). 

Groen, Andrew. 2015. \textit{Empires of EVE: A History of the Great Wars of EVE Online}.

Hafner-Burton, Emilie M., Miles Kahler, and Alexander H. Montgomery. 2009. ``Network Analysis for International Relations." \textit{International Organization}, 63, pp. 559-92.

Huremovic, Kenan. 2014. ``Rent Seeking and Power Hierarchies: A Noncooperative Model of Network Formation with Antagonistic Links." FEEM Working Paper No. 45.2014.

Jackson, Michael O. 2008. \textit{Social and Economic Networks}. Princeton University Press.

Jervis, Robert. 1994. ``Hans Morgenthau, Realism, and the Scientific Study of International Politics." \textit{Social Research}, Vol. 61, No. 4. 

Kaplan, Morton A. 1957. \textit{System and Process in International Politics}.
ECPR Press.

Keohane, Robert O. (ed.). 1986. \textit{Neorealism and Its Critics}. Columbia University Press.

Machiavelli, Niccolò and Peter Constantine. 2007.  \textit{The Essential Writings of Machiavelli}. Modern Library.

Mearsheimer, John J. 2014. \textit{The Tragedy of Great Power Politics}. W. W. Norton \& Company.

Morgenthau, Hans. 1954. \textit{Politics Among Nations: The Struggle for Power and Peace}. Knopf.

Poulshock, Michael. 2017. ``An Abstract Model of Historical Processes." \textit{Cliodynamics: The Journal of Quantitative History and Cultural Evolution}, issue 8(1). 

Powell, Robert. 1991. ``Absolute and Relative Gains in International Relations Theory." \textit{The American Political Science Review}, Vol. 85, No. 4, pp. 1303-1320.

Schroeder, Paul. 1994. ``Historical Reality vs. Neo-Realist Theory." \textit{International Security}, Vol. 19, No. 1, pp. 108-148.

Sheehan, Michael. 1995. \textit{Balance of Power: History \& Theory}. Routledge.

Stoll, Richard J. 1987. ``System and State, in International Politics: A Computer Simulation of Balancing in an Anarchic World." \textit{International Studies Quarterly}, Vol. 31, 387-402.

Tellis, Ashley J. 1995. ``Reconstructing Political Realism: The Long March to Scientific Theory." \textit{Security Studies}, Vol. 5, No. 2, pp. 3-94.

Thucydides. 1954. \textit{History of the Peloponnesian War}. Penguin Classics.

Vasquez, John A. and Colin Elman. 2003. \textit{Realism and the Balancing of Power: A New Debate}. Prentice Hall.

Waltz, Kenneth. 1979. \textit{Theory of International Politics}. McGraw-Hill.

Wohlforth, William C. (et al). 2007. ``Testing Balance-of-Power Theory in World History." \textit{European Journal of International Relations}, Vol. 13(2), pp. 155-185.

\end{hangparas}
\end{footnotesize}

\end{document}